\documentstyle[preprint,aps]{revtex}
\begin{document}
\tightenlines
\title{All-optical switching with bacteriorhodopsin}   
\author{Sukhdev Roy and Parag Sharma}
\address{Department of Physics and Computer Science, Dayalbagh Educational Institute, Agra 282 005, India.}
\author{Aditya K. Dharmadhikari and Deepak Mathur}
\address{Tata Institute of Fundamental Research, 1 Homi Bhabha Road, Mumbai 400 005, India.}
\date{\today}
\maketitle
\begin{abstract}
All-optical, mirrorless switching and bistability is demonstrated with bacteriorhodopsin (bR). A low-power, 532 nm laser beam modulates the transmission of a cw laser beam at 635 nm that corresponds to peak absorption of the O-excited state in the bR photocycle. The switch has features like a NOT-gate; its switching contrast depends on the pulse width and average power of the modulating laser. The phase of the switching characteristics can be controlled by varying the pulse width and frequency of the modulating laser. Simulations based on a rate equation approach considering a six-state model of the bR photocycle successfully reproduce the experimental results.    
\end{abstract}
 
A switch is the basic building block of information processing systems. Recent years have witnessed tremendous research interest to develop all-optical switching and routing for high speed, high-bandwidth communication and computing \cite{bishop}. The key element in an all-optical switch is a material that exhibits large nonlinear optical response. An ideal switching material would also be expected to be small in size and weight, and possess high intrinsic speed, low propagation delay and power dissipation, and offer the possibility of tailoring properties for device applications \cite{henari}. Bacteriorhodopsin (bR) is perhaps the simplest of nature's photon-driven molecular machines that offers tantalizing possibilities of optical processing at a molecular level \cite{forchel}; it exhibits high quantum efficiency, large absorption cross-section and nonlinearities as well as stability against environmental perturbations. A natural photochromic protein, bR is found in the purple membrane of Halobacterium halobium and has emerged as a prime candidate for bio-molecular photonic applications.  By absorbing green-yellow light, wild-type bR protein undergoes several structural transformations in a well-characterized photocycle, B$_{570}\rightarrow$K$_{610}\rightarrow$L$_{540}\rightarrow$M$_{410}\rightarrow$N$_{550}\rightarrow$O$_{640}\rightarrow$B$_{570}$, where letters denote the different electronic states, and subscripts correspond to their respective peak absorption wavelengths (in nm).  All-optical switching in bR has been reported earlier using complex, multiple-laser geometries involving holograms \cite{holograms}, refractive index modulation \cite{omos} and enhanced photo-induced anisotropy \cite{wu}. In this Letter, we demonstrate an extremely simple, mirrorless all-optical switch with bR using a low-power cw ``signal" beam at 635 nm that is switched by a pulsed ``modulating" beam at 532 nm. Our switch exhibits bistability and exhibits features like a NOT-gate. The pulse width and intensity of the modulating beam affects the switching characteristics and this is further probed by theoretical simulations using a rate equation approach. 

Fig. 1 shows a schematic of the simple experimental set up. The film of wild-type bR used in our experiments had a pH of 7, and optical density 3 at 570 nm. The 532 nm modulating beam was obtained from a frequency-doubled Nd:YVO4 laser. A mechanical chopper was used to change the modulation frequency. The signal beam from a 10.8 mW cw 635 nm diode laser, was switched by absorption in the bR film. The wavelength of 635 nm is in close proximity to the peak absorption wavelength (640 nm) of the intermediate O-state in the bR photocycle. The diameters of the modulating and signal beams were 4 mm and 2 mm, respectively. These beams were monitored by photodiodes connected to a digital oscilloscope. 
The performance of the switch was studied over a range of pulse widths (3-16 ms) and power (5-58 mW) of the modulating beam. By way of illustration, experimental manifestation of all-optical switching is shown in Fig. 2(a) for a modulating pulse width of 3.3 ms and average incident power of 39 mW; the modulating pulse profile and the temporal variation of transmitted signal intensity (TSI) is shown for wild-type bR. The TSI is initially high (switch``on" state) due to relatively low linear absorption. When the modulating laser beam irradiates the sample, it activates the bR photocycle and, hence, excites bR molecules such that the O-state becomes populated. This leads to increased absorption of the signal beam and consequent decrease in its transmission (switch ``off" state). In our series of experiments we have observed that as the modulating pulse width increases, the percentage modulation of the transmitted signal beam also increases.  

Theoretical simulations have been carried out by considering bR molecules exposed to two light beams of intensities $I_m$ (modulating) and $I_s$ (signal), that modulate the population densities ($N$) of different states through  excitation and de-excitation processes. They are described by rate equations of the form $dN/dt = \hat{O}N$, where the $\hat{O}$ operator is defined in terms of photo-induced and thermal transitions of all the six states (B, K, L, M, N and O) in the bR photocycle using methodology that has been described in detail elsewhere \cite{roy}. Excitation of the bR sample has been considered due to both the 532 nm and 635 nm laser beams, as the signal beam is not very weak in comparison to the modulating beam. The propagation of the signal beam is governed by $I_s = I_{s0} exp [-\alpha_s(I_m,I_s) L]$, where $\alpha_s(I_m,I_s)$ is the intensity-dependent absorption coefficient of the signal beam written here as $\alpha_s(I_m,I_s) = N_B(I_m,I_s)\sigma_{Bs} + N_K(I_m,I_s)\sigma_{Ks} + N_L(I_m,I_s)\sigma_{Ls} + N_N(I_m,I_s)\sigma_{Ns} + N_O(I_m,I_s)\sigma_{Os}$,  where $\sigma$ is the absorption cross-section of the state denoted by the subscript and $s$ denotes its value at the wavelength of the signal beam. The experimental curves have been modeled by a super-Gaussian modulating laser pulse of the form $I_m = I_{m0} exp [-2^{2m} ln2 ((t-t_0)/\Delta t)^{2m}]$, where $m$ is the pulse profile parameter and $\Delta t$ is the pulse width. The quantum efficiency for transitions B$\leftrightarrow$K and the film thickness ($L$), have been considered to be 0.64 and 60 mm respectively. The theoretical simulations have been carried out using typical values of absorption cross-sections and rate constants of intermediate states \cite{forchel,ludman}.

The simulated optical switching curves (Fig. 2b) were obtained for the same conditions as the experimental measurement (Fig. 2a), namely, $\Delta t$ = 3.3 ms, $n$ = 208.3 Hz, $m$ = 2 at fixed average power of 39 mW. Both the switch on/off time and percentage modulation was observed to increase with pulse width (Dt), with the modulation level saturating after a certain value. For instance, for $\Delta t$ = 3.3 ms, the measured ``off" and ``on" times are 3.0 ms and 1.8 ms respectively, in reasonable accord with corresponding simulated values of 2.8 ms and 2.0 ms. The symmetry of the switching curves expectedly increases for Dt greater than the relaxation time of the complete bR photocycle, as this conforms to the steady state case. 
Switching characteristics measured with somewhat higher average modulating power (58 mW) are shown in Fig. 3a, for $\Delta t$ = 16 ms. The corresponding simulated curves are shown in Fig. 3b with $m$ = 5. As before, an increase in the average modulating power results in an increase in the percentage modulation, which saturates at higher values. From Figs. 2 and 3, it is evident that simulated results are in good accord with experimental results even though the asymmetric profile of the input modulating pulse has been approximated by a symmetric super-Gaussian function.
 
The optical bistability that is a feature of our switch is depicted in Fig. 3 where the functional relationship between the modulating and output signals is, characteristically, in the form of a loop in each of the two cases shown (for different values of $\Delta t$). Bistabilities in optical switches have been studied for more than 25 years, almost always in the context of complex schemes that employ linear or non-linear interferometric geometries \cite{bistabilities}. In such geometries, changes in the light intensity within the ring cavity lead to variations in the refractive index of the non-linear medium.  This means that the phase change experience by the light when travelling through such medium is altered; this continues until a stable point is reached and results in bistable behaviour.  In the case of bR, however, the bistability that we observe is attributed to the intensity-dependent absorption coefficient, $\alpha_s(I_m,I_s)$. The non-square nature of the loops that are shown in Fig. 3 is a consequence of the six different values of $\alpha$ that are involved in the bR photocycle, each of them being intensity dependent. This leads to a ``washing out" of the system response. Non-square loops have been observed in earlier studies \cite{rao} employing single-wavelength excitation. Unlike previous work, the orientation of the loop that we observe indicates that our switch exhibits features that are akin to a NOT-gate; this is consistent with the phase lag between input and output signals that are observed in our data, including the subset that is shown in Figs. 2 and 3. We found little evidence of ringing or over-shooting, features that frequently manifest themselves in bistable operation of cavity-based schemes \cite{bistabilities}.  

Our simulations have enabled some further, important, insights to be obtained. We find that decrease in values of rate coefficients $k_M$, $k_N$, and $k_O$ increases the switch ``off" and ``on" time that, in turn, results in a shift of the switching curves towards longer time with respect to the minima of the modulating pulses. The switching characteristics are more sensitive to variation in $k_N$ and $k_O$ than to $k_M$. Moreover, decrease in values of $k_N$ and $k_O$ lowers percentage modulation contrary to earlier indications \cite{roy} from studies carried out for a single modulating pulse, in which the percentage modulation was found to increase. The present work differs significantly in that we now probe the effect of $k_M$ and $k_O$ on the switching characteristics for a train of pulses. We find that as the first modulating pulse arrives, the population of the O-state increases, giving rise to increased absorption of the signal beam and consequent switching ``off" of the TSI. The arrival of the next modulating pulse in the train stops the signal beam from rising to its initial value, and decreases the TSI. After a few successive modulating pulses, the switching contrast remains constant and appears as shown in Figs. 2 and 3. Increase in $k_O$ results in an increase in the symmetry of the switching characteristics as the total relaxation time of the photocycle becomes less than the modulating pulse width and conforms to the steady state case. Variation in $k_O$ also results in a shift in the average power of the transmitted signal beam, which increases with $k_O$.

We find that the switching characteristics are also sensitive to the profile of the modulating pulse. Decreasing $m$ makes the input pulse profile nearly Gaussian, and the switching characteristics become more symmetric due to gradual variation in intensity with time, which is of the order of the lifecycle. Increasing m initially makes the pulse rise sharply and results in an increase in the switch ``off" time and decrease in the switch ``on" time that, after a certain value of $m$, saturates. The peaks in the signal transmission also shift to longer time with respect to the minima of the modulating pulses, as the pulse on-off time is unequal. The percentage modulation of the TSI also increases and saturates at higher $m$-values and the peak-to-peak variation in intensity of the normalized input pulses becomes nearly unity. For the same film, decreasing the modulating pulse width and increasing its frequency results in decrease in the percentage modulation and a shift in the switching characteristics such that the modulation of the output signal beam appears to be in phase with the input modulating pulses. For instance, for a peak modulating intensity of 468.6 mW cm$^{-2}$, modulating pulse width of $\Delta t$ = 0.15 ms, and a pulse train frequency of 3.3 kHz, the TSI becomes in phase with the modulating pulse train. This is an important result as one can control the phase of the switching characteristics by varying the modulating pulse width and its frequency.

The percentage modulation can be increased by selecting the modulating beam at 570 nm, which corresponds to the peak absorption of the initial B state. Moreover, the percentage modulation is very sensitive to the absorption of the signal beam by the initial B state. Hence, narrowing the B-state absorption spectrum such that it does not absorb the signal beam ($\sigma_{Bs}$ = 0), can lead to higher switching contrast (the signal beam can be nearly completely switched ``off" by the modulating beam). For instance, in the present case, for $\Delta t$ = 16 ms at $P_{av}$ = 58 mW, for the same parameters, the percentage modulation would increase from 14\% to 85\% as $\sigma_{Bs}\rightarrow$ 0. Decreasing the photocycle time with $\sigma_{Bs}$ = 0, further increases the percentage modulation. Hence, complete switching at low powers and low switching time can be achieved when the signal beam is not absorbed by the initial B-state ($\sigma_{Bs}$ = 0), and when the total photocycle time is small. Adjusting the modulating beam intensity also controls the switching contrast. Since the molecular properties of bR, such as the relaxation rates, absorption cross-sections and spectra of the intermediate states are amenable to modification by physical, chemical and genetic engineering techniques, the switching characteristics of bR can be tailored for specific applications \cite{forchel}. Since bR can also be processed as a large aperture film, or in a crystalline form for 2D/3D applications, bR-based all-optical switches would be potentially useful in applications that involve provisioning of light paths in optical cross-connects and add-drop systems in wavelength division multiplexing optical networks. Controlling the phase of the switching characteristics by the modulating pulse width and frequency may also prove to be of utility in future designs of all-optical light modulators and phase shifters.

In summary, we have shown absorptive, all-optical switching and bistability with bR, with advantages of small size, extremely simple low-power operation, small linear absorption coefficient, mirror-less structure and flexibility of design. A simulation analysis has also been made and theoretical switching characteristics have been shown to be in good agreement with experimental results.

SD is grateful to All-India Council for Technical Education and the Department of Science and Technology, and PS to the Council for Scientific and Industrial Research for partial support of this work.

\begin{figure}
\caption{Experimental set-up for all-optical, mirrorless switching in bR film.} 
\end{figure}

\begin{figure}
\caption{Measured (a) and simulated (b) transmitted signal intensity at 635 nm (dashed line) and input modulating beam at 532 nm (solid line)  for pump pulse width $\Delta t$ = 3.3 ms; $n$ = 208.3 Hz, $m$ = 2 with average power of 39 mW. The simulated data are normalized to the corresponding measured signals.}
\end{figure}

\begin{figure}
\caption{Measured (a) and simulated (b) transmitted signal intensity at 635 nm (dotted line), and input modulating beam at 532 nm (solid line) with 58 mW average pump power, for $\Delta t$ = 16 ms, $n$ = 40.9 Hz and $m$ = 5. The simulated data are normalized to the corresponding measured signals.}
\end{figure} 

\begin{figure}
\caption{Variation of transmitted signal intensity with input modulating signal for (a) $\Delta t$ = 3.3 ms and (b) $\Delta t$ = 16 ms.}
\end{figure}  

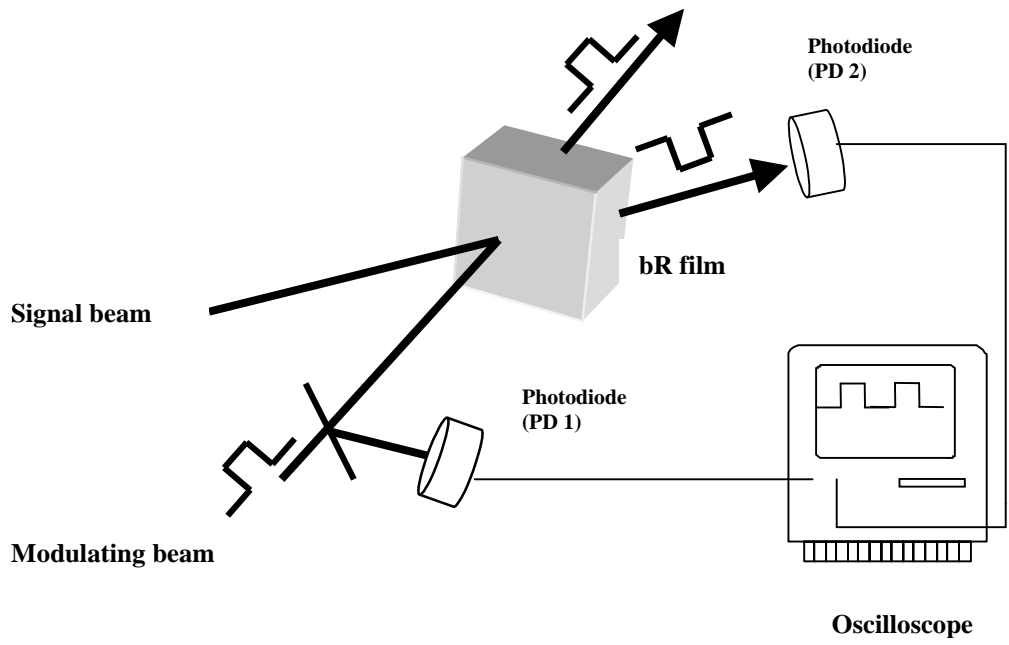
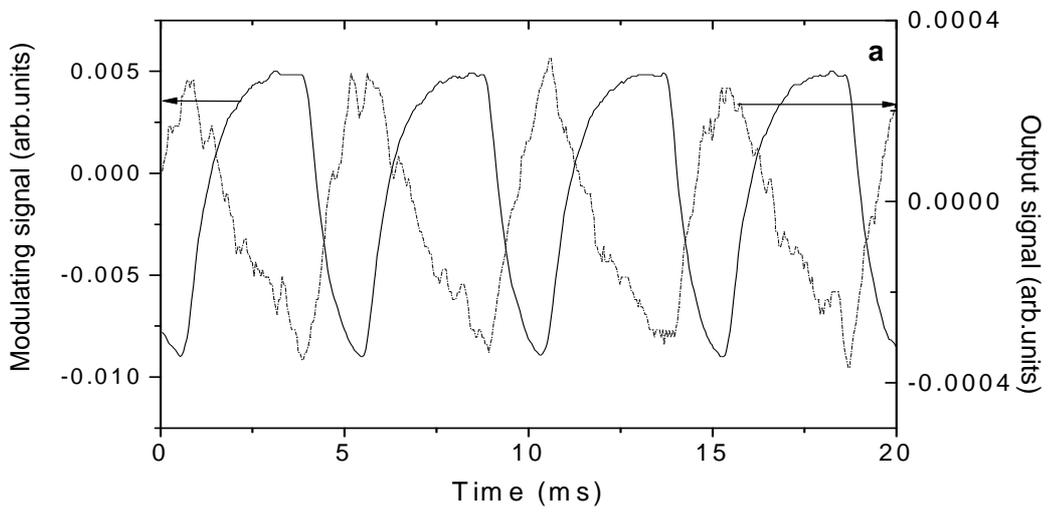
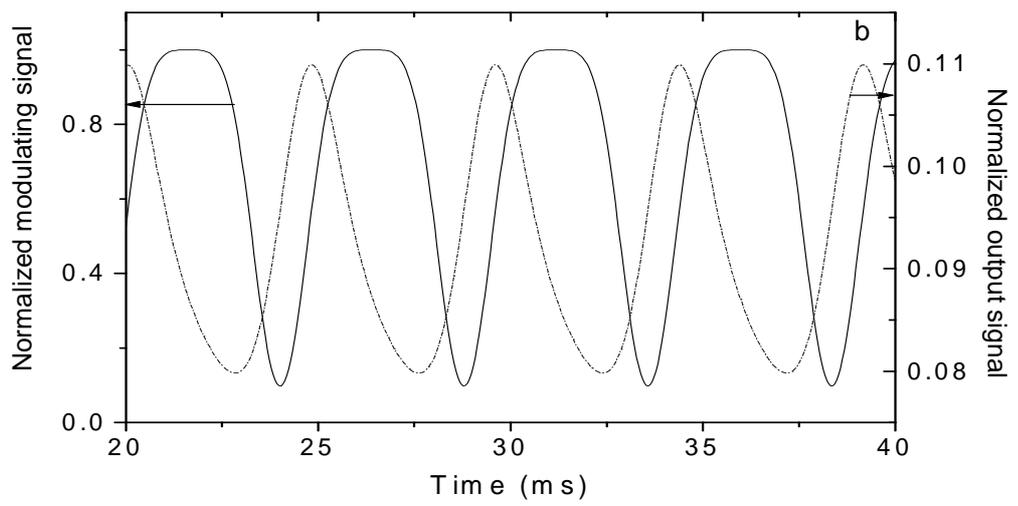
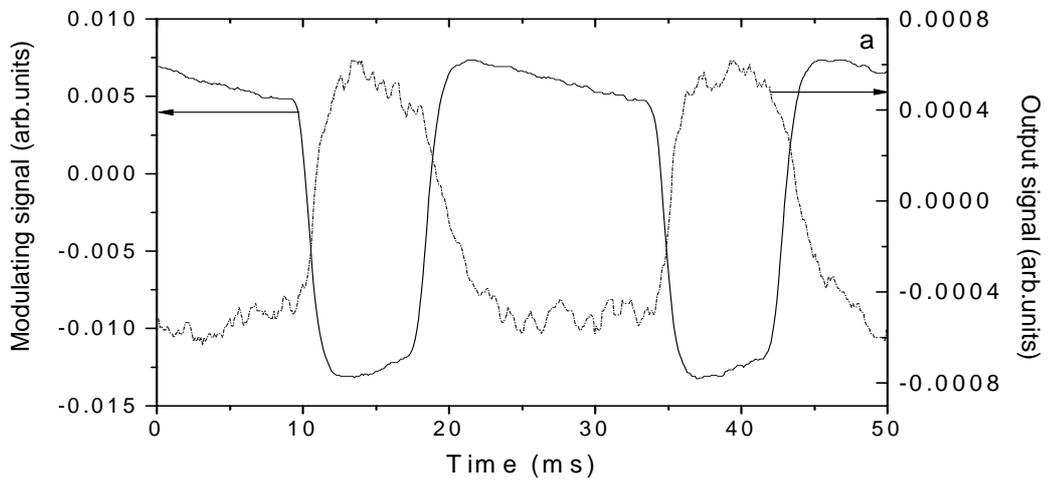
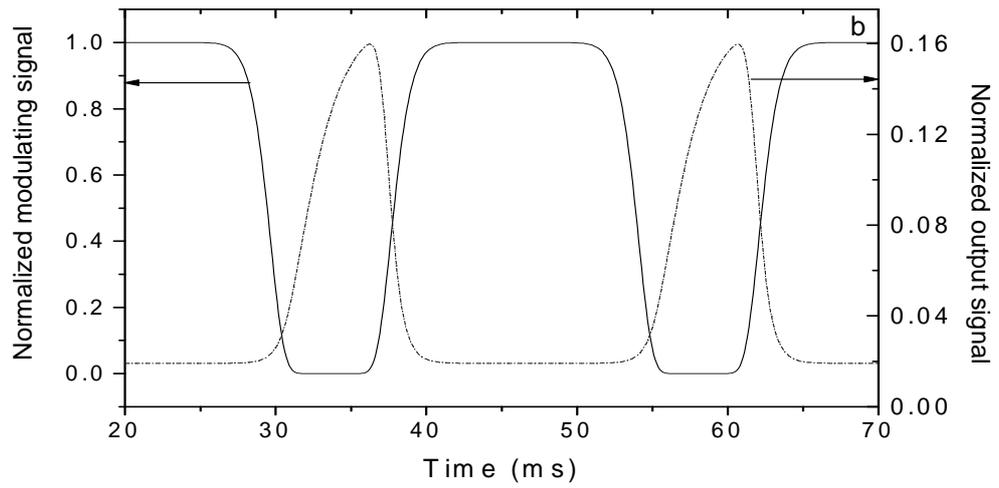
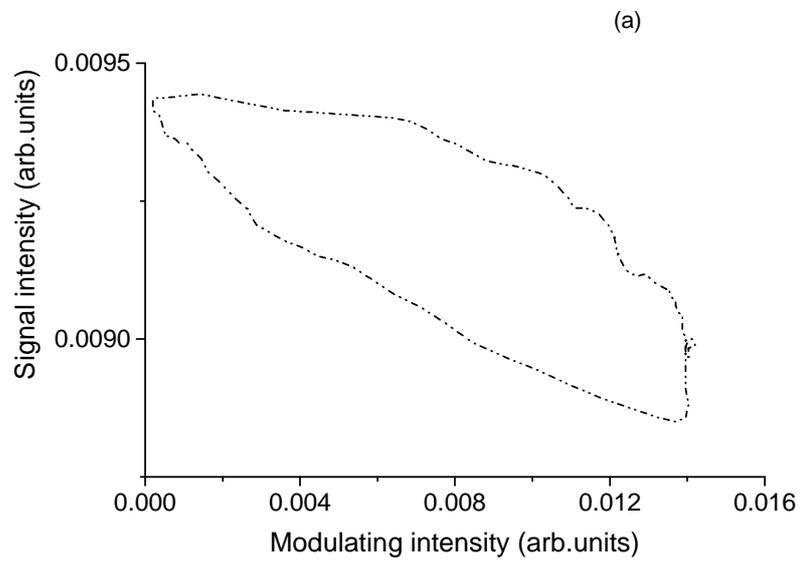
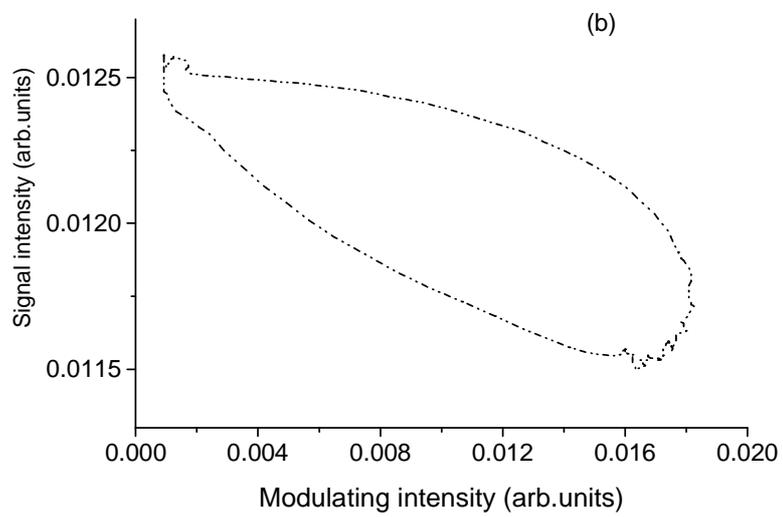
\end{document}